\documentclass[10pt, twocolumn, aps, prd, amssymb, amsmath, superscriptaddress, eqsecnum, nofootinbib, notitlepage, tightenlines, longbibliography]{revtex4-1}

\usepackage{braket,verbatim}
\usepackage{graphicx,tikz,bbm,bm}
\usepackage[colorlinks=true, citecolor=cyan, linkcolor=., anchorcolor=.,filecolor=.,menucolor=.,runcolor=.,urlcolor=.,]{hyperref}

\newcommand{\be}{\begin{equation}}
\newcommand{\ee}{\end{equation}}

\newcommand{\mcH}{\mathcal{H}}
\newcommand{\vg}{\vec{g}}

\newcommand{\Tr}{\text{Tr}}
\newcommand{\mcO}{\mathcal{O}}
\newcommand{\mfh}{\mathfrak{h}}
\newcommand{\mfH}{\mathfrak{H}}
\newcommand{\mcN}{\mathcal{N}}
\newcommand{\mcP}{\mathcal{P}}

\def\eqref#1{(\textcolor{blue}{\ref{#1}})}

\begin{document}

\title{Thermal representations in group field theory: \\ squeezed vacua and quantum gravity condensates}

\author{Mehdi Assanioussi}
\email{mehdi.assanioussi@desy.de}
\affiliation{II. Institute for Theoretical Physics, University of Hamburg, Luruper Chaussee 149, 22761 Hamburg, Germany}

\author{Isha Kotecha}
\email{isha.kotecha@aei.mpg.de}
\affiliation{Max Planck Institute for Gravitational Physics (Albert Einstein Institute), Am M\"{u}hlenberg 1, 14476 Potsdam-Golm, Germany \\ }
\affiliation{Institut f\"{u}r Physik, Humboldt-Universit\"{a}t zu Berlin, Newtonstra{\ss}e 15, 12489 Berlin, Germany}

\begin{abstract}

We apply the formalism of thermofield dynamics to group field theory quantum gravity and construct thermal representations associated with generalised equilibrium Gibbs states using Bogoliubov transformations. The newly constructed class of thermal vacua are entangled, two-mode squeezed, thermofield double states. The corresponding finite temperature representations are inequivalent to the standard zero temperature one based on a degenerate vacuum. An interesting class of states, coherent thermal states, are defined and understood as thermal quantum gravity condensates.

\end{abstract}

\maketitle


\section{Introduction}

An increasing number of studies are hinting toward intimate links between entanglement and geometry. Spacetime is thought to be highly entangled, with quantum correlations in the underlying quantum gravitational system being crucial. Particularly in discrete quantum gravity approaches, a geometric phase of the universe is expected to emerge from a quantum, pre-geometric one via a phase transition. This geometric phase must then also be highly entangled. Squeezing techniques have been used often in such contexts to define suitably entangled non-perturbative vacua. Also, the emergent phase must encode a sufficient notion of semi-classicality. This notion is commonly introduced via coherent states. Moreover, fluctuations in relevant observables are expected to be important in the physics of a quantum spacetime and thus must be included in the description of the system. In this work we show how to construct quantum gravitational phases in group field theory (for now, kinematically) using thermofield dynamics, which display all these features, namely entanglement, coherence and statistical fluctuations in given observables. 

Such phases can be of significant relevance in the study of semi-classical and effective continuum descriptions of quantum spacetime, especially in the context of cosmology and black holes. For instance, in the context of the present work, a semi-classical approximation resides in considering a coherent state, with its characteristic single-particle wavefunction being the relevant dynamical collective variable, while a continuum approximation resides in considering a non-perturbative (inequivalent) condensate phase of the quantum gravity system, with an infinite number of underlying quanta. Having said that, in general, identifying suitable semi-classical and continuum limits in discrete quantum gravity is a known open issue, and we do not address it directly in this work.

Group field theories (GFT) \cite{Freidel:2005qe,Oriti:2006se,Oriti:2011jm} are statistical field theories of combinatorial and algebraic quanta of geometry \cite{Oriti:2013aqa,Chirco:2018fns,Kotecha:2019vvn}, formally defined by a partition function 
\be Z_{\rm GFT} = \int [D\varphi_\kappa] \; e^{-S(\{\varphi_\kappa\})} \nonumber \,. \ee
They are strictly related to various other approaches like loop quantum gravity \cite{Ashtekar:2004eh, thiemann_2007, Han:2005km, Bodendorfer:2016uat}, spin foams \cite{Perez:2012wv, rovelli_vidotto_2014}, causal dynamical triangulations \cite{Loll:2019rdj}, tensor models \cite{Gurau:2011xp} and lattice quantum gravity \cite{Hamber:2009mt}. Like in usual field theories, the kinematics is specified by a choice of the fields $\varphi_\kappa$, each defined in general over a domain space of direct products of Lie groups, and taking values in some target vector space. The dynamics is specified by propagators and interaction vertices encoded in a function $S$, which can be understood as a Landau-Ginzburg free energy function in the present statistical context \cite{Kotecha:2019vvn}; or as a Euclidean action from the point of view of standard quantum field theories \cite{Reisenberger:2000zc,Freidel:2005qe,Oriti:2006se,Oriti:2011jm}. However unlike in usual field theories, $S$ is non-local in general with respect to the base manifold. This non-locality is essential and encodes the non-trivial combinatorial nature of the fundamental degrees of freedom and their dynamics. Moreover, the base manifold is not spacetime, but carries algebraic information associated with discrete geometric and matter degrees of freedom. Such a complete absence of any continuum spacetime structures a priori is a manifestation of background independence in group field theory, like in various other non-perturbative approaches to quantum gravity. 

The partition function $Z_{\rm GFT}$ perturbatively generates Feynman diagrams that are labelled 2-complexes (dual to labelled stranded diagrams), with boundary states given by labelled graphs \cite{Reisenberger:2000zc,Freidel:2005qe,Oriti:2006se,Oriti:2011jm}. For the choice of models closer to loop quantum gravity and spin foam setups, the boundary states are abstract spin networks (but organised in a second quantised Hilbert space, that of a field theory \cite{Oriti:2013aqa}) and bulk processes are spin foams, both of which in turn are dual to polyhedral complexes when restricting to loopless combinatorics \cite{Bianchi:2010gc,Oriti:2014yla}. Thus, a group field theory generates discrete quantum spacetimes made of fundamental polyhedral quanta\footnote{A quantised polyhedron with $d$ faces is dual to a gauge-invariant open $d$-valent spin network node \cite{Bianchi:2010gc}. The latter is in fact a special case (namely, a $d$-patch \cite{Oriti:2014yla}) of richer combinatorial boundary structures which can be treated analogously in our present setup, but in that case without any related discrete geometric understanding of the same. For more details, we refer to discussions in \cite{Kotecha:2019vvn,Oriti:2014yla} and references therein.}. \

We can describe the same structures from a many-body perspective \cite{Oriti:2017twl}, and treat as more fundamental an interacting system of many such quanta. This viewpoint enables us to import formal techniques from standard many-body physics for macroscopic systems, by treating a quantum polyhedron or an open spin network node as a single atom or particle of interest \cite{Kotecha:2018gof,Chirco:2018fns,Kotecha:2019vvn,Chirco:2019kez}. This further leads to a modelling of an extended region of discrete quantum space (a spin network) as a multi-particle state with a large number of entangled quanta (for instance, via gluing conditions) \cite{Bianchi:2012ev,Chirco:2017xjb,Baytas:2018wjd}; and a region of dynamical quantum spacetime (a spin foam) as an interaction process. This is the perspective that we employ here to successfully use techniques from thermofield dynamics, even when working with a radically different kind of system, one that is background independent, and devoid of any standard notion of space, time and other associated geometric structures and standard matter couplings. 

Thermofield dynamics (TFD) is an operator framework for finite temperature quantum field theory \cite{Takahasi:1974zn,Matsumoto:1985mx,Umezawa:1982nv,Umezawa:1993yq,Khanna:2009zz}. One of its main advantages lies in the fact that its formulation parallels that of zero temperature quantum field theory, thus powerful tools of the temperature-independent setup can be translated to the thermal case, including perturbative Feynman diagrammatic techniques, symmetry breaking analyses, and for what concerns us here, Fock space techniques. TFD has been applied in various fields such as superconductivity, quantum optics, and string theory. 

The core idea of the TFD formalism is to represent statistical ensemble averages as temperature-dependent vacuum expectation values. That is, given a system described by a Hilbert space $\mathfrak{h}$ and an algebra of observables $\mathfrak{A}_\mathfrak{h}$ on $\mathfrak{h}$, one looks for a vector state $\ket{\Omega_\rho}$ (thermal vacuum) in a Hilbert space $\mathfrak{H}$, corresponding to a density operator ${\rho}$ on $\mathfrak{h}$, such that the following condition holds for all observables $A$ of the system,
\be \label{req} \Tr_{\mathfrak{h}}( {\rho}  {A}_{\mathfrak{h}}) = \bra{\Omega_\rho} {A}_{\mathfrak{H}}\ket{\Omega_\rho}_{\mathfrak{H}}  \ee
where the subscript $\mathfrak{h} \; \rm or \; \mathfrak{H}$ denotes a suitable representation of a classical observable in the respective Hilbert spaces. In this work, we are mainly concerned with statistical equilibrium, thus with density operators of the Gibbs form $e^{-\beta  {O_\mathfrak{h}}}$. Note that we will not make use of hats to denote operators in order to declutter the overall notation in the following.  \

A vector state satisfying equation \eqref{req} can only be defined in an extended Hilbert space, by adding the so-called tilde degrees of freedom to the original ones \cite{Takahasi:1974zn}. Importantly, this doubling, or in general an enlargement of the space of the relevant degrees of freedom, is a characteristic feature of finite temperature description of physical systems. This was also discovered in algebraic quantum field theory for equilibrium statistical mechanics \cite{Haag:1967sg}. Moreover, equation \eqref{req} is strongly reminiscent of the construction of a GNS representation \cite{Bratteli:1996xq} induced by an algebraic statistical state, with the vector state $\ket{\Omega_\rho}$ being the cyclic vacuum of a new thermal representation. These intuitions have in fact led to tangible relations between the two formalisms, with the tilde degrees of freedom of TFD being understood as those of the conjugate representation of Tomita-Takesaki theory in the algebraic framework \cite{Ojima:1981ma,Matsumoto:1985mx,Landsman:1986uw,Celeghini:1998sy,Khanna:2009zz}. 

These structures are also encountered commonly in quantum information theory \cite{nielsen_chuang_2010}, which in turn is utilised heavily in various areas of modern theoretical physics, like holography. Specifically, constructing a state $\ket{\Omega_\rho}$ is nothing but a purification of $ {\rho}$. A prime example of a vector state is the thermofield double state, which is the purification of a Gibbs density operator. These states are used extensively in studies probing connections between geometry, entanglement, and more recently, complexity \cite{VanRaamsdonk:2016exw,Chapman:2018hou}. For instance in AdS/CFT, this state is important because a bulk with an eternal AdS black hole is dual to a thermofield double in the boundary quantum theory \cite{Maldacena:2001kr}. 

Our goal is to construct entangled, thermofield double states $\ket{\Omega_\rho}$ that encode statistical fluctuations\footnote{The vector state $\ket{\Omega_\rho}$ essentially encodes the density operator $\rho$, as reflected by equation \eqref{req}. Thus, like in standard treatments of systems in state $\rho$, statistical fluctuations in state $\ket{\Omega_\rho}$ can be investigated for instance in terms of variances of relevant observables of the system.} with respect to observables of interest, along with their associated inequivalent representations; and to subsequently introduce a family of coherent configurations of the quanta of geometry built upon the thermal vacuum.
The use of such thermal quantum gravity condensates, say with respect to a spatial volume observable in cosmology \cite{KA,Oriti:2016qtz,Oriti:2016acw,Gielen:2016dss,Pithis:2019tvp}, can be understood as implementing both semi-classical and continuum approximations. This is in the precise sense of working with a non-perturbative coherent condensate phase. The novel feature of the coherent thermal phase though is that of statistical fluctuations associated with the thermal vacuum, in addition to quantum fluctuations inherent in any quantum state. Statistical fluctuations in general are inevitable in macroscopic systems and could be non-trivial in both the semi-classical and continuum limits. This is unlike purely quantum fluctuations which are expected to be negligible in a semi-classical limit, but of significance especially at early times \cite{Gielen:2017eco}.

The article is organised as follows. We begin with an overview of the relevant essentials of the formalism of thermofield dynamics in section \ref{tfd}, and use them in the rest of the paper to present a systematic extension of group field theories to construct finite temperature phases associated with generalised equilibrium Gibbs states. In section \ref{alg} we give a brief overview of the original setup of a bosonic group field theory coupled to a scalar matter field, which provides the basis for a zero temperature phase as described in section \ref{zero}. In section \ref{volgibbs}, we briefly recall the subtleties surrounding the foundational issue of defining statistical equilibrium in background independent systems to consider generalised Gibbs states, while focusing on extensive observables as generators. 
We construct their corresponding thermal vacua (thermofield double states), along with the inequivalent quantum gravity phases that they generate in section \ref{finite}. In section \ref{CTS} we introduce the class of coherent thermal states and briefly overview their properties. We conclude with an outlook on future prospects for further applications in group field theory and other quantum gravity approaches in section \ref{conclusion}. 


\section{Thermofield dynamics} \label{tfd}

Below we review the basics of the TFD formalism \cite{Takahasi:1974zn,Matsumoto:1985mx,Umezawa:1982nv,Umezawa:1993yq,Khanna:2009zz} using a simple example of an oscillator. We will extend it to a field theory setup directly for GFTs in the subsequent sections. \

Consider a single bosonic oscillator, described by ladder operators $ {a}, {a}^\dag$ satisfying the commutation algebra,
\begin{gather}
 [ {a}, {a}^\dag] = 1 \ ,\ [ {a}, {a}]  = [ {a}^\dag, {a}^\dag]= 0
\end{gather}
with the $a$-particle vacuum being specified by ${a}\ket{0} = 0$. A Fock space $\mfh$ is generated by actions of polynomial functions of the ladder operators on $\ket{0}$. Thermal effects are then encoded in density operators defined on $\mfh$. In particular, an equilibrium state at inverse temperature $\beta$ is a Gibbs state,
\be \label{qmrho}  {\rho}_\beta = \frac{1}{Z} e^{-\beta  {H}} \ee
where $ {H}$ is a Hamiltonian operator on $\mfh$, possibly of grand-canonical type. But we must note that in the context of background independent systems with an absence of a notion of absolute energy, one may work with other observables like the spatial volume, as will be discussed briefly in section \ref{volgibbs} \cite{Kotecha:2018gof,Kotecha:2019vvn}. \

This system is extended by including tilde degrees of freedom, with a Hilbert space $\tilde{\mfh}$ generated by a vacuum $\ket{\tilde{0}}$ and the ladder operators $ {\tilde{a}}, {\tilde{a}}^\dag$ satisfying also the bosonic algebra,
\be [ {\tilde{a}}, {\tilde{a}}^\dag] = 1 \ee
with $[ {\tilde{a}}, {\tilde{a}}] = [ {\tilde{a}}^\dag, {\tilde{a}}^\dag] = 0$, and ${\tilde{a}}\ket{\tilde{0}} = 0$. All tilde and non-tilde degrees of freedom commute with each other, that is 
\be [ a,  {\tilde{a}}]=[  {a},  {\tilde{a}}^\dag]= [  {a}^\dag,  {\tilde{a}}]=[  {a}^\dag,  {\tilde{a}}^\dag]=0 \;. \ee
The Hilbert space $\tilde{\mfh}$ is conjugate to the original one via the action of the modular conjugation operator of KMS theory \cite{Ojima:1981ma}; or equivalently via the tilde conjugation rules of TFD: $i.\, (AB)\tilde{\,} = \tilde{A}\tilde{B} ;\ ii. \, (A^\dag)\tilde{\,} = \tilde{A}^\dag ;\ iii. \,  (\tilde{A})\tilde{\,} = A ;\ iv. \,  (z_1 A + z_2 B)\tilde{\,} = \bar{z}_1 \tilde{A} + \bar{z}_2\tilde{B} ;\ v. \, \ket{0}\tilde{\,} = \ket{\tilde{0}}$, for all non-tilde and tilde operators defined on $\mfh$ and $\tilde{\mfh}$ respectively, and $z_i \in \mathbb{C}$. 

The zero temperature (or its inverse, $\beta=\infty$) phase of the system is described by the enlarged Hilbert space,
\be \mfH_\infty = \mfh \otimes \tilde{\mfh} \ee
built from the Fock vacuum,
\be \ket{0_\infty} = \ket{0} \otimes \ket{\tilde{0}} \ee
by actions of the ladder operators $a,a^\dag,\tilde{a},\tilde{a}^\dag$, such that
\be a\ket{0_\infty} = \tilde{a}\ket{0_\infty} = 0 \;.\ee

Temperature is introduced via thermal Bogoliubov transformations of the algebra generators, \be\{a,a^\dag,\tilde{a},\tilde{a}^\dag\}_{\beta = \infty} \mapsto \{b,b^\dag,\tilde{b},\tilde{b}^\dag\}_{0 < \beta < \infty}\ee given by
\begin{align} \label{qmbog1}
b &= \cosh \theta(\beta) \, a - \sinh \theta(\beta) \, \tilde{a}^\dag  \\ \label{qmbog2}
\tilde{b} &= \cosh \theta(\beta) \, \tilde{a} - \sinh \theta(\beta) \, {a}^\dag 
\end{align}
along with expressions for their adjoints $b^\dag$ and $\tilde{b}^\dag$. The Bogoliubov transformations are canonical, thus leaving the algebra unchanged, so that the $\beta$-ladder operators also satisfy bosonic commutations relations,
\begin{align} \label{qmbcr1}
[ b, b^\dag] = [ \tilde{b}, \tilde{b}^\dag] = 1 \\ \label{qmbcr2}
[ b, \tilde{b} ]=[ b,  \tilde{b}^\dag]=0
\end{align}
along with their adjoints. The temperature-dependent annihilation operators now specify a thermal vacuum,
\be \label{qmbvac} b\ket{0_\beta} = \tilde{b}\ket{0_\beta} = 0  \ee
which is cyclic for the thermal Hilbert space $\mfH_\beta$. It is an example of a two-mode squeezed state, which can be restored to its full generality most directly by adding a net phase difference between the $\cosh$ and $\sinh$ terms in the transformations \eqref{qmbog1} and \eqref{qmbog2}. \

The Hilbert space $\mfH_\beta$ can most directly be organised as a Fock space with respect to the $\beta$-dependent ladder operators that create and annihilate $b$-quanta over the thermal background $\ket{0_\beta}$. Like in any other Fock space construction, one can define useful classes of states in $\mfH_\beta$. We will return to this point in section \ref{CTS} where we define one such interesting class of states in the quantum gravity system, namely the coherent thermal states. \

The thermal Bogoliubov transformations \eqref{qmbog1} are parametrized by $\theta(\beta)$, which must thus encode complete information about the corresponding statistical state. In the present case, it must be uniquely associated with the Gibbs state $\rho_\beta$, which has a well known characteristic Bose number distribution,
\be \Tr_{\mfh}(\rho_\beta a^\dag a) = \frac{1}{e^{\beta \omega}-1} \ee
using a Hamiltonian of the form $H = \omega a^\dag a$, and the number operator $N=a^\dag a$ of the physical, non-tilde system. This is how $\theta$ is usually determined in TFD, using equation \eqref{req} for the number operator. Then, using inverse Bogoliubov transformations, the right hand side of equation \eqref{req} for the present case gives
\be \bra{0_\beta}a^\dag a \ket{0_\beta}_{\mfH_\beta} = \sinh^2\theta(\beta) \;.  \ee
This specifies $\theta$ via the equation,
\be \frac{1}{e^{\beta \omega}-1} = \sinh^2\theta(\beta) \;. \ee

The non-tilde degrees of freedom can be understood as being physically relevant in the sense that they describe the subsystem of interest which is accessible to the observer. In other words, it is the subsystem under study in any given situation. Then, the physically relevant observables belong to the algebra restricted to the non-tilde degrees of freedom, which can be retrieved from the full description by partially tracing away the complement, here the tilde degrees of freedom. Thus, one is usually interested in observable averages of the form $\bra{0_\beta} \mcO(a,a^\dag) \ket{0_\beta}$, for operators $\mcO$ that are in general polynomial functions of the generators of the physical non-tilde algebra. Notable geometric examples include relativistic quantum field vacua in Minkowski and Schwarzschild spacetimes, where the non-tilde algebra has support on spacetime regions exterior to the respective horizons. Here, the tilde subsystems are CPT conjugates (modulo a rotation) of the non-tilde ones, and belong to the interior of the horizons \cite{Israel:1976ur,Kay:1985zs,Sewell:1982zz}. Another physical interpretation of the tilde subsystem that is more common in condensed matter theory, is that of a thermal reservoir \cite{Umezawa:1982nv,Strocchi:2008gsa}. \

In the present discrete quantum gravity context, for now we retain the elementary, quantum information-theoretic interpretation of the non-tilde and tilde degrees of freedom, simply as describing a given subsystem and its complement respectively, without assigning any further geometric meanings. \

The two sets of ladder operators are related to each other explicitly via equations \eqref{qmbog1} and \eqref{qmbog2}. This suggests that their respective vacua are also related by an associated transformation. Indeed they are, via the following unitary transformation\footnote{The form of this unitary operator, along with equation \eqref{qmuvac}, shows that the thermal vacuum $\ket{0_\beta}$ is a two-mode squeezed state in which $a\tilde{a}$-pairs have condensed \cite{Umezawa:1993yq}.}
\be \label{qmu} U(\theta) = e^{\theta(\beta)(a^\dag \tilde{a}^\dag - a \tilde{a})} \;. \ee
In terms of this thermal squeezing operator, we then have
\be \label{qmuvac} \ket{0_\beta} = U(\theta)\ket{0_\infty} \ee
and
\begin{align}  b &= U(\theta) \, a\, U(\theta)^{-1} \\ \label{qmub}
 \tilde{b} &= U(\theta) \, \tilde{a}\, U(\theta)^{-1} \;. \end{align}

It is clear then that such unitary operators map the different $\beta$-representations into each other. In finite quantum systems this is simply a manifestation of von Neumann's uniqueness theorem. However, when extending the above setup to quantum field theory, one would expect that representations at different temperatures are inequivalent. This is certainly the case for physical systems in general. It is also true in the present quantum gravity case, as we show in section \ref{finite}. Lastly, we note that in the field theory extension, equations \eqref{qmbog1}-\eqref{qmbvac} hold mode-wise and still remain well-defined. Together, they describe the $\beta$-phase of the system. However, the operator $U(\theta)$ is no longer well-defined in general (before any cut-offs). Thus, without any suitable regularization, equations \eqref{qmu}-\eqref{qmub} technically do not hold in full field theory.


\section{Bosonic group field theory} \label{alg}

The most commonly studied models in GFT are scalar field theories defined over multiple copies of the local gauge group of gravity, which is $SL(2,\mathbb{C})$ for 4-dim Lorentzian models and $Spin(4)$ for Euclidean ones. The subgroup $SU(2)$ is chosen often for boundary configurations related to loop quantum gravity. For models of pure geometry (that is, before any matter coupling), the group field $\varphi$ is then commonly taken to be a complex-valued scalar field on $SU(2)^d$.
This choice of $\varphi$ corresponds to taking an atom of space to be a geometric polyhedron with $d$ faces (under an additional closure condition, see below), or equivalently, an open $d$-valent node with each of its incident half-link labelled with an $SU(2)$ element. \

In this work, we consider the scalar field of the theory to be defined over the manifold $SU(2)^d \times \mathbb{R}$. From a technical point of view, this allows a more general treatment in the formalism, where the base manifold has a non-compact direction. From a physical point of view, this can be viewed as coupling gravity to a scalar matter field, which can be used to define a relational frame of reference \cite{Li:2017uao,Oriti:2016qtz}, crucial for our subsequent application to thermal condensate cosmology \cite{KA}. Thus, the group field under consideration\footnote{The techniques and results of this work can be straightforwardly extended to include multiple fields and multiples copies of $\mathbb{R}$ in the domain manifold.} here is
\be  \varphi : SU(2)^d \times \mathbb{R} \to \mathbb{C} : \vg,\phi \mapsto \varphi(\vg,\phi) \,. \ee
In the following, we will use the notations $\vg \equiv g_i$ $(i=1,\dots,d)$ interchangeably as per convenience. The field $\varphi$ is further chosen to be invariant\footnote{The right gauge invariance is imposed in order to avail a quantum geometric interpretation of the quanta of the GFT field. An additional left gauge invariance can also be imposed in the context of homogeneous cosmologies \cite{Oriti:2016qtz,Gielen:2016dss}. However, our setup will technically follow through with or without (either or both) these additional symmetries and their associated geometric meanings.} under a diagonal right action of $SU(2)$ on $SU(2)^d$, that is
\be \label{ginv} \varphi(g_i,\phi) = \varphi(g_i h, \phi) \;\;\;\;\;\; \forall h \in SU(2) \;. \ee
This is the geometric condition of closure, which in the corresponding quantised theory leads to the understanding of a quantum of this field as a quantised polyhedron, in turn dual to a gauge-invariant spin network node \cite{Barbieri:1997ks,Baez:1999tk,Bianchi:2010gc}. This invariance effectively reduces the geometric part of the domain space to $SU(2)^d/SU(2)$. In the following however, we will continue to use a redundant parametrisation for convenience and consider the gauge-invariant functions to be defined on full $SU(2)^d$ while explicitly satisfying equation \eqref{ginv}. \

The corresponding quantum operator theory is based on a commutations relations algebra for bosonic\footnote{The zero and finite temperature formulations discussed here can be extended to fermionic statistics with an anti-commutations relations algebra.} polyhedra,
\be \label{occr} [  {\varphi}(\vg,\phi),  {\varphi}^\dag(\vg',\phi')] = \mathbb{I}(\vg,\vg')\delta(\phi-\phi') \ee
with $[ {\varphi}, {\varphi}]=[ {\varphi}^\dag, {\varphi}^\dag]=0$, and $\mathbb{I}$ being a delta distribution on $SU(2)^d$ compatible with gauge invariance, given by
 \be \mathbb{I}(\vg,\vg') = \int_{SU(2)}dh \; \prod_{i=1}^d \delta(g_i h g_i'^{-1})\;. \ee
 
We note that this quantum theory is in fact based on an abstract Weyl algebra for GFT \cite{Kegeles:2017ems}. Our constructions here to access an inequivalent thermal phase using TFD may in principle also be formulated more rigorously using modular structures of Tomita-Takesaki theory and their relation to TFD \cite{Landsman:1986uw,Ojima:1981ma,Celeghini:1998sy}. In the present work however, we are content with a more physical approach to the problem, directly in line with usual TFD studies.


\section{Degenerate vacuum and Zero temperature phase} \label{zero}

The zero temperature phase of the system is based on an enlargement of the Fock representation of the above bosonic (Weyl) algebra, along the lines presented in section \ref{tfd} but generalised here to a field theory. \

The Hilbert space for a single gauge-invariant quantum is the state space of geometries of a quantum polyhedron with an additional real degree of freedom,
\begin{align} \mcH &=  L^2(SU(2)^d/SU(2) \times \mathbb{R}) \nonumber \\
 &\cong L^2(SU(2)^d/SU(2)) \otimes L^2(\mathbb{R}) \end{align}
where the quotient by $SU(2)$ ensures gauge invariance. In order to work with formally well-defined quantities, we smear the operator-valued distributions $ {\varphi}(\vg,\phi)$ with a suitable basis of functions in $\mcH$.\footnote{For more functional rigour, the test functions may be defined on the dense subspace of smooth functions $C^\infty(.) \subset L^2(.)$. But this technicality is often overlooked in practice, as is also done here, thus taking the $L^2$ space to be the space of single particle wave functions and the space of test functions.} For gauge-invariant functions on $SU(2)^d$, a useful basis is given by a set of gauge-invariant Wigner functions 
\be \label{wigner} D_{J}(\vg) = \sum_{\vec{n}}C^{\vec{j}}_{\vec{n} \iota} \prod_{i=1}^d D^{j_i}_{m_in_i}(g_i) \;\;,\;\;J \equiv (\vec{j},\vec{m},\iota)   \ee 
where $j_i  \in \mathbb{N}/2$ labels irreducible representations of $SU(2)$, $m_i,n_i \in (-j_i,...,+j_i)$ are matrix indices in representation $j_i$, $D^{j_i}_{m_in_i}$ are complex-valued Wigner matrix coefficients (multiplied by a factor of $ \sqrt{2j_i+1}$ for normalisation) in representation $j_i$, and $C^{\vec{j}}_{\vec{n} \iota}$ are intertwiner basis elements indexed by $\iota$ arising due to the closure condition in \eqref{ginv}. Orthonormality and completeness are respectively given by,
\begin{align} \int d\vg \; \bar{D}_{J}(\vg) D_{J'}(\vg) &= \delta_{JJ'} \;, \\
\sum_{J} \bar{D}_{J}(\vg) D_{J}(\vg') &= \mathbb{I}(\vg,\vg') \;.  \end{align}

Similarly for the matter part, let us consider a basis of complex-valued smooth functions $T_\alpha(\phi)$ in $L^2(\mathbb{R})$, labelled by a discrete index $\alpha$, satisfying orthonormality and completeness,
\begin{align} \int d\phi \; \bar{T}_\alpha(\phi) T_{\alpha'}(\phi) &= \delta_{\alpha \alpha'} \;,\\
 \sum_\alpha \bar{T}_\alpha(\phi) T_\alpha(\phi') &= \delta(\phi-\phi') \;.\end{align}
We thus have a complete orthonormal basis on $\mcH$ consisting of functions $f_{J\alpha}$ of the tensor product form,
\be f_{J\alpha}(\vg,\phi) = (D_J \otimes T_\alpha)(\vg,\phi) = D_J(\vg)T_\alpha(\phi) \,. \ee
Then a suitable set of mode ladder operators can be defined by smearing the operators $\varphi , \varphi^\dag$ with this basis,
\begin{align} \label{a}  
 {a}_{J\alpha} :=   {\varphi}(f_{J\alpha}) &= \int_{SU(2)^d \times \mathbb{R}} d\vg d\phi\; \bar{D}_J(\vg)\bar{T}_\alpha(\phi)   {\varphi}(\vg,\phi) \;,\\
 \label{adag}   {a}^\dag_{J\alpha} =   {\varphi}^\dag(f_{J\alpha}) &= \int_{SU(2)^d \times \mathbb{R}} d\vg d\phi\; {D}_J(\vg) {T}_\alpha(\phi)   {\varphi}^\dag(\vg,\phi) \,. 
 \end{align}
This essentially decomposes the operators $  {\varphi},  {\varphi}^\dag$ in terms of the modes $f_{J\alpha}$, which can be seen directly by inverting the above two equations. The algebra relations are,
\be \label{RegCom} [  {a}_{J\alpha},  {a}^\dag_{J'\alpha'}] = \delta_{JJ'} \delta_{\alpha \alpha'} \ee
and $[  {a},  {a}]=[a^\dag,a^\dag]=0$. Note that, thanks to the choice of a discrete basis $\{T_\alpha(\phi)\}_\alpha$ in $L^2(\mathbb{R})$, the algebra commutator in \eqref{RegCom} produces Kronecker deltas, instead of the Dirac delta distribution present in the original basis in equation \eqref{occr}. The consideration of a regular algebra instead of a distributional one is an important feature in the present work, particularly for the $\phi$-modes, where we deal with inequivalent representations obtained via Bogoliubov transformations. Namely, the Kronecker delta $\delta_{\alpha\alpha'}$ is crucial in order to avoid divergences related to the coincidence limit of $\delta(\phi-\phi')$. Such terms with $\delta(\phi-\phi')$ arise naturally when calculating thermal expectation values of certain relevant observables, for example the average thermal number density in $\phi$-basis (compare with equation \eqref{thnum}).

The vacuum is specified by, 
\be \label{degvac} {a}_{J\alpha}\ket{0} = 0 \;\;\; \forall J,\alpha \ee 
which is a degenerate state with no discrete geometric or matter data, dubbed often as a no-space state. It generates the symmetric Fock space
\be \mcH_F = \bigoplus_{\mcN \geq 0} {\rm sym}\, \mcH^{\otimes \mcN} \ee
by cyclic action of the generators $\{  {a}_{J\alpha},   {a}^{\dag}_{J\alpha},  {1}\}$ of this representation\footnote{This is the GNS representation of the GFT Weyl algebra induced by the algebraic Fock state \cite{Kegeles:2017ems,Kotecha:2018gof}.}. For instance, a single particle ($\mcN=1$), single mode state is,
\be \ket{J\alpha} \equiv \ket{f_{J\alpha}} =   {a}^\dag_{J\alpha} \ket{0} \ee
while a generic single particle state with a wavefunction $\psi(\vg,\phi) = \sum_{J,\alpha} \psi_{J\alpha}f_{J\alpha}(\vg,\phi) \in \mcH$ is
\be \ket{\psi} =   {a}^\dag(\psi)\ket{0} = \sum_{J,\alpha}\psi_{J\alpha}  {a}^\dag_{J\alpha}\ket{0} \,. \ee

The zero temperature phase is then given by extending the above with the conjugate representation space $\tilde{\mcH}_F$, as discussed in section \ref{tfd}. This gives the zero temperature ($\beta = \infty$) description in terms of a Hilbert space,
\be \mcH_\infty = \mcH_F \otimes \tilde{\mcH}_F \ee
which is a Fock space on the cyclic vacuum 
\be \ket{0_\infty} = \ket{0} \otimes \ket{\tilde{0}} \ee
with ladder operators $ \{ a, {a}^\dag, \tilde{a},  {\tilde{a}}^\dag \}_{J,\alpha}$ that satisfy,
\begin{align}
[  {a}_{J\alpha},  {a}^\dag_{J'\alpha'}] &= \delta_{JJ'} \delta_{\alpha \alpha'} \\
[  {\tilde{a}}_{J\alpha},  {\tilde{a}}^\dag_{J'\alpha'}] &= \delta_{JJ'} \delta_{\alpha \alpha'}
\end{align}
and $[  {a},  {a}]=[  {\tilde{a}},  {\tilde{a}}]=[  {a},  {\tilde{a}}]=[  {a},  {\tilde{a}}^\dag]=0$. The non-tilde operators describing the subsystem of interest are those defined in \eqref{a} and \eqref{adag}, while the tilde operators of the complement are
\begin{align}   {\tilde{a}}_{J\alpha} &= \int_{SU(2)^d \times \mathbb{R}} d\vg d\phi\; {D}_J(\vg){T}_\alpha(\phi)  {\tilde{\varphi}}(\vg,\phi) \;, \\
   {\tilde{a}}^\dag_{J\alpha} &= \int_{SU(2)^d \times \mathbb{R}} d\vg d\phi\; \bar{D}_J(\vg) \bar{T}_\alpha(\phi)  {\tilde{\varphi}}^\dag(\vg,\phi) \;. \end{align}
The vacuum satisfies
\be    {a}_{J\alpha}\ket{0_\infty} =    {\tilde{a}}_{J\alpha}\ket{0_\infty} = 0 \;\;\; \forall J,\alpha \;. \ee
The action of all polynomial functions of non-tilde and tilde ladder operators on $\ket{0_\infty}$ generates $\mcH_\infty$, all in complete analogy with standard $\mcH_F$, including the construction of multi-particle states, coherent states, squeezed states, and so on. 


\section{Generalised Gibbs states} \label{volgibbs}

The familiar way to include thermal effects is with statistical states, as density operators in a given representation or in general as algebraic mixed states. As we discussed in section \ref{tfd}, an equivalent way is with their corresponding vector states (thermal vacua) in an enlarged representation of the system, in which the additional degrees of freedom are known to be integral for encoding finite temperature effects. \

Here we are interested in Gibbs density operators for describing equilibrium phases of the quantum gravity system. But defining statistical equilibrium in background independent settings, like the present one, is a subtle issue (see \cite{Kotecha:2019vvn}, and references therein, for detailed discussions). For the purposes of this work however, what concerns us is the so-called thermodynamical characterisation, based on the maximum entropy principle, for defining generalised Gibbs states of the form
\be \rho_{\{\beta_\ell\}} = \frac{1}{Z_{\{\beta_\ell\}}}e^{-\sum_{\ell}\beta_\ell \mcO_\ell} \ee
where $\beta_\ell$ are generalised inverse temperatures conjugate to a given set of observables $\mcO_\ell$ \cite{Kotecha:2018gof,Chirco:2018fns,Kotecha:2019vvn}. This state is a result of maximising the information entropy, $-\langle \ln \rho \rangle_\rho$, under the set of constraints $\langle \mcO_\ell \rangle_{\rho} = \rm constant$ and $\langle 1 \rangle_{\rho} = 1$. Particularly, observables $\mcO$ need not be dynamical energies. This is especially valuable in background independent systems where the notion of energy is ambiguous at best, or not defined at all in more radical setups such as the current quantum gravitational one. In fact, these observables can be geometric operators such as area and volume, which could be particularly relevant in quantum gravity
\footnote{We note that classical quantities such as space(-time) volume, or the area of a closed surface, are not necessarily well defined in a diffeomorphism invariant (background independent) context. This is because such quantities are at most diffeomorphism covariant, thus failing to represent a physical gauge-invariant observable. This open issue of observables in general relativity, arises also within any background independent quantum gravity approach. Having said that, one could still construct well defined quantum operators, which can be formally associated to certain, classically peculiar, quantities in a given quantum framework, such as the volume of space(-time). In this sense, one could still have operators with a geometric interpretation in a background independent context, like in GFT or in loop quantum gravity.}.
For instance, in the context of applying these constructions to homogeneous cosmology, $\mcO$ could be chosen to be the spatial volume, thus giving a volume Gibbs state \cite{Kotecha:2018gof}. Such a state would then encode thermal fluctuations in spatial volume of the underlying discrete quantum space, which may be expected to be important in cosmological dynamics, especially at very early times \cite{KA}.\

Below we consider the more general case with self-adjoint and semi-bounded operators for group field theories coupled with a scalar matter, so that for such generalised Gibbs states, the above machinery of TFD can be applied to construct various different phases characterised by the corresponding thermofield doubles. Naturally, this leaves open the possibilities of different observables, and which ones are relevant in any given situation is also an important part of the broader problem of investigating the statistical mechanics of quantum gravity for an emergent, thermodynamical spacetime with features compatible with semi-classical studies \cite{Kotecha:2018gof,Kotecha:2019vvn}. \

Given a self-adjoint and semi-bounded $\mcO$, and using the maximum entropy procedure recalled above, under the constraints $\langle  {\mcO} \rangle_\rho = \rm constant$, $\langle  {N} \rangle_\rho = \rm constant$ and $\langle  1 \rangle_\rho = 1$, a generalised Gibbs state can be defined as,
\be  \label{rho} {\rho}_{\beta,\mu} = \frac{1}{Z_{\beta,\mu}} e^{-\beta( {\mcO} - \mu  {N})} \ee
where $\beta \in \mathbb{R}$ and $\mu \in \mathbb{R}$ are chosen such that the combination $\beta( {\mcO} - \mu  {N})$ is a positive operator, thus ensuring proper normalisation of the state. 
A particularly interesting class of observables where the partition function can be evaluated rather straightforwardly is for positive, extensive operators. That is, we can consider the class of Gibbs states characterised by a self-adjoint, positive and extensive operator on the original system $\mcH_F$ given by,
\be \label{vol}  {\mcP} = \sum_{J,\alpha} \lambda_{J\alpha}  {a}^\dag_{J\alpha} {a}_{J\alpha}\ee
with $\forall J,\alpha, \; \lambda_{J,\alpha} \in \mathbb{R}_+$. Extensive operators in GFT are a second quantisation of those loop quantum gravity operators which are diagonal in some intertwiner basis, such as the spatial volume operator \cite{Ashtekar:1997fb, Oriti:2013aqa, Kotecha:2018gof}.

Extensive operators are compatible with the total number operator,
\be  {N} = \sum_{J,\alpha}  {a}^\dag_{J\alpha} {a}_{J\alpha} \;, \ee 
that is, $[\mcP,N]=0$. They are thus diagonal in an occupation number basis of $\mcH_F$ consisting of multi-particle, multi-mode states of the general form,
\be \label{occ} \ket{\{n_{J\alpha}\}} = \frac{1}{\sqrt{n_{J_1\alpha_1}! n_{J_2\alpha_2}!...}}  {a}_{J_1\alpha_1}^{\dag \, n_{J_1\alpha_1}}  {a}_{J_2\alpha_2}^{\dag \, n_{J_2\alpha_2}}... \ket{0} \ee
which are orthonormal,
\be \langle \{n_{J\alpha}\} | \{m_{J'\alpha'}\} \rangle = \delta_{n_{J\alpha}m_{J'\alpha'}}\delta_{JJ'}\delta_{\alpha\alpha'} \;. \ee

The partition function for \eqref{rho} with operators $\mcP$ can then be evaluated\footnote{Given a suitable set of properly normalised modes $f_{J\alpha}$ as done in \ref{zero} above, the calculation for this partition function is along the lines of that showed for a volume Gibbs state in \cite{Kotecha:2018gof}, to which we refer for details.} in the basis of states \eqref{occ} to give,
\be Z_{\beta,\mu} = \prod_{J,\alpha} \frac{1}{1-e^{-\beta (\lambda_{J\alpha} - \mu)}}  \ee
where $\mu < \min (\lambda_{J\alpha})$ for all $J$ and $\alpha$. This is a grand-canonical state of $a$-particles, which essentially describes a gas of these atoms of space with a changing total number in the given (non-tilde) subsystem. Technically, the number operator in \eqref{rho} simply implements a constant shift by $\mu$ in the observable spectrum, which for now can be neglected by fixing $\mu$ to an arbitrary value, or by simply replacing $\lambda_{J\alpha} - \mu \mapsto \lambda_{J\alpha}$. \

It is evident that by construction, the parameter $\beta$ controls the strength of statistical fluctuations in $\mcP$, regardless of any other interpretations. Note however, that one can reasonably inquire about its geometric meaning, especially if the operator $\mcP$, which is its thermodynamic conjugate, has a clear geometric interpretation. It would thus be interesting to investigate this aspect in a concrete example where the choice of the observable is adapted to a physical context, like cosmology (see for instance \cite{KA}). \

Finally, the ensemble average for number density will be useful for the construction of the associated thermal representation as outlined in section \ref{tfd}. For the state above, it is given by the characteristic Bose distribution,
\be \label{num} \Tr_{\mcH_F}(\rho_{\beta, \mu} a^\dag_{J\alpha}a_{J\alpha}) = \frac{1}{e^{\beta (\lambda_{J\alpha}-\mu)}-1} \;, \ee
from which the average total number $\braket{N}$ of polyhedral quanta can be obtained by summing over all $J$ and $\alpha$. Partial sums over either $J$ or $\alpha$ would give average number densities $\braket{N_\alpha}$ or $\braket{N_J}$ respectively. In the context of relational dynamics, for instance in GFT cosmology \cite{Oriti:2016qtz,Oriti:2016acw,Gielen:2016dss,Pithis:2019tvp}, quantities like $\braket{N_\alpha}$ are strictly related to relational observables as functions of the matter variable $\phi$, the details of which are included in \cite{KA}.


\section{Thermal squeezed vacuum and Finite temperature phase} \label{finite}

Now that we have chosen a suitable class of thermal states of equation \eqref{rho} characterised with operators $\mcP$, we can proceed to define its associated thermal phase generated by a thermal vacuum and $\beta$-dependent ladder operators $\{b_{J\alpha},b^\dag_{J\alpha},\tilde{b}_{J\alpha},\tilde{b}^\dag_{J\alpha}\}_\beta$, along the lines detailed in section \ref{tfd}. \

Thermal Bogoliubov transformations\footnote{More general two-mode squeezing transformations can also be considered by taking a net phase difference between the two contributions, say $e^{i\delta}$ scaling the $\sinh$ terms.} give the new ladder operators, mode-wise,
\begin{align} \label{bog1}
b_{J\alpha} &= \cosh \theta_{J\alpha}(\beta) \, a_{J\alpha} - \sinh \theta_{J\alpha}(\beta) \, \tilde{a}_{J\alpha}^\dag  \\ \label{bog2}
\tilde{b}_{J\alpha} &= \cosh \theta_{J\alpha}(\beta) \, \tilde{a}_{J\alpha} - \sinh \theta_{J\alpha}(\beta) \, {a}_{J\alpha}^\dag 
\end{align}
along with their adjoints $b_{J\alpha}^\dag$ and $\tilde{b}_{J\alpha}^\dag$. Inverse transformations are,
\begin{align} \label{invbog1}
a_{J\alpha} &= \cosh \theta_{J\alpha}(\beta) \, b_{J\alpha} + \sinh \theta_{J\alpha}(\beta) \, \tilde{b}_{J\alpha}^\dag  \\ \label{invbog2}
\tilde{a}_{J\alpha} &= \cosh \theta_{J\alpha}(\beta) \, \tilde{b}_{J\alpha} + \sinh \theta_{J\alpha}(\beta) \, {b}_{J\alpha}^\dag 
\end{align}
and their respective adjoints. The $\beta$-dependent annihilation operators specify the thermal vacuum via
\be b_{J\alpha}\ket{0_\beta} = \tilde{b}_{J\alpha}\ket{0_\beta}  = 0 \;, \ee
thus giving the finite temperature Hilbert space $\mcH_\beta$. $\ket{0_\beta}$ is a concrete example of a (class of) thermofield double state(s) in discrete quantum gravity. It is an entangled state encoding quantum correlations between pairs of $a_{J\alpha}$ and $\tilde{a}_{J\alpha}$ polyhedral quanta. Further, using equations \eqref{req}, \eqref{num}, and
\be \label{thnum} \bra{0_\beta}a_{J\alpha}^\dag a_{J\alpha}\ket{0_\beta}_{\mcH_\beta} = \sinh^2 \theta_{J\alpha}(\beta) \;,\ee
the parameters $\theta_{J\alpha}$ can be determined from 
\be \label{num2} \sinh^2\theta_{J\alpha}(\beta) = \frac{1}{e^{\beta \lambda_{J\alpha}}-1} \;.  \ee
Note that the singular case in equation \eqref{num2}, or equivalently in \eqref{num}, can be understood as Bose-Einstein condensation to the ground state of $\mathcal{P}$ (equation \eqref{vol}) in the present thermal gas of quantum gravitational atoms. In the context of a volume Gibbs state, such a phenomenon was first observed in \cite{Kotecha:2018gof}, and was used to show a model-independent, purely statistical mechanism for the emergence of a low-spin phase.

Lastly, the $\beta$-phase that we have constructed here, being described kinematically by $ \{\ket{0_\beta}, b_{J\alpha},b^\dag_{J\alpha},\tilde{b}_{J\alpha},\tilde{b}^\dag_{J\alpha} \}$, is inequivalent to the zero temperature phase $\{  \ket{0_\infty}, a_{J\alpha},a^\dag_{J\alpha},\tilde{a}_{J\alpha},\tilde{a}^\dag_{J\alpha} \}$. This can be seen directly from the transformation equations between the two vacua:
\begin{align} \ket{0_\beta} &= U(\theta)\ket{0_\infty} \\ &= e^{\sum_{J,\alpha}\theta_{J\alpha}(a_{J\alpha}^\dag\tilde{a}_{J\alpha}^\dag - a_{J\alpha}\tilde{a}_{J\alpha})} \ket{0_\infty} \\
&= e^{-\sum_{J,\alpha} \ln \cosh\theta_{J\alpha}}e^{\sum_{J,\alpha}a^{\dag}_{J\alpha}\tilde{a}^\dag_{J\alpha}\tanh \theta_{J\alpha}} \ket{0_\infty} \\
&= \prod_{J,\alpha} \frac{1}{ \cosh\theta_{J\alpha} } \times e^{\sum_{J,\alpha}a^{\dag}_{J\alpha}\tilde{a}^\dag_{J\alpha}\tanh \theta_{J\alpha}} \ket{0_\infty}  \end{align}
Clearly, the pre-factor of the product of inverse $\cosh$ functions vanishes in general, without any cut-offs in the modes. This means that the overlap between the two vacua is zero, and the two representations built upon them are inequivalent. In other words, this transformation in field theory is ill-defined in general due to an infinite number of degrees of freedom, giving rise to inequivalent representations describing distinct phases of the system.


\section{Coherent thermal states} \label{CTS}

We now define a family of coherent states in the thermal representation, called coherent thermal states. We understand them as defining thermal quantum gravity condensates, expected to be relevant in the studies of semi-classical and continuum approximations in discrete quantum gravity models based on polyhedral quanta of geometry. 
Indeed, unlike the purely thermal state $\ket{0_\beta}$, coherent states can encode a notion of semi-classicality, with which one can attempt to extract effective dynamics from an underlying quantum gravity model. For instance in group field theory, it has been shown that a coherent condensate phase of the $a$-quanta can support FLRW cosmological dynamics, and thus represents a viable choice of a quantum gravitational phase relevant in the cosmological sector \cite{Oriti:2016qtz,Oriti:2016acw,Gielen:2016dss,Pithis:2019tvp,Kegeles:2017ems}. But a coherent state over the degenerate vacuum \eqref{degvac} is unentangled, and we expect a geometric phase of the universe to be highly entangled. Moreover these states cannot in themselves encode statistical fluctuations in different observable quantities. Therefore, from a physical point of view, the construction of coherent thermal states is amply justified.

Coherent thermal states \cite{1985JOSAB...2..467B,MANN1989273,mann,Khanna:2009zz} are a coherent configuration of quanta over the thermal vacuum, implemented by displacing $\ket{0_\beta}$ with displacement operators of the form,
\be \label{dis} D_a(\sigma) = e^{a^\dag(\sigma) - a(\sigma)}  \ee
for $\sigma \in \mcH$. To recall, the usual coherent states $\ket{\sigma} \in \mcH_F$ of $a$-particles are,
\be \ket{\sigma} := D_a(\sigma) \ket{0} \;, \ee
while $\ket{\tilde{\sigma}} \in \tilde{\mcH}_F$ for $\tilde{a}$-particles are,
\be \ket{\tilde{\sigma}} := D_{\tilde{a}}(\sigma) \ket{\tilde{0}} \;. \ee
The tilde in the ket notation $\ket{\tilde{\sigma}}$ simply means that the state is an element of the conjugate Hilbert space, and $D_{\tilde{a}}$ is a displacement operator of the same form as \eqref{dis} but for tilde ladder operators.    \

The most useful property of these states is that they are eigenstates of their respective annihilation operators,
\begin{align}
a_{J\alpha}\ket{\sigma} = \sigma_{J\alpha}\ket{\sigma} \\
\tilde{a}_{J\alpha}\ket{\tilde{\sigma}} = \sigma_{J\alpha}\ket{\tilde{\sigma}}
\end{align}
which is at the heart of their extensive use as robust, most classical-like, quantum states.

Notice that under the tilde conjugation rules stated in section \ref{tfd}, we have
\be (\ket{\sigma} \otimes \ket{\tilde{0}})\tilde{\,} = \ket{0} \otimes \ket{\tilde{\bar{\sigma}}} \ee
in $\mcH_\infty$. That is, coherent states $\ket{\sigma} \in \mcH_F$ and $\ket{\bar{\sigma}} \in \tilde{\mcH}_F$, are conjugates of each other. In other words, the following state
\be \ket{\sigma,\bar{\sigma};\infty} \equiv \ket{\sigma} \otimes \ket{\tilde{\bar{\sigma}}} = D_a(\sigma)D_{\tilde{a}}(\bar{\sigma})\ket{0_\infty} \in \mcH_\infty \ee
of the full system at zero temperature is self-conjugate,
\be \ket{\sigma,\bar{\sigma};\infty}\tilde{\,} = \ket{\sigma,\bar{\sigma};\infty}\;. \ee

In the finite temperature phase then, coherent thermal states \cite{1985JOSAB...2..467B,MANN1989273,mann,Khanna:2009zz} are defined as the following self-conjugate states encoding coherence in the original $a$ degrees of freedom over the thermal vacuum,
\be \label{cts} \ket{\sigma,\bar{\sigma};\beta} := D_a(\sigma)D_{\tilde{a}}(\bar{\sigma})\ket{0_\beta} \in \mcH_\beta \;. \ee
Being elements of $\mcH_\beta$, as expected they are eigenstates of the $\beta$-annihilation operators $b_{J\alpha}$ with temperature-dependent eigenfunctions,
\begin{align}
b_{J\alpha}\ket{\sigma,\bar{\sigma};\beta} = (\cosh\theta_{J\alpha} - \sinh\theta_{J\alpha})\sigma_{J\alpha}\ket{\sigma,\bar{\sigma};\beta} ,\\
\tilde{b}_{J\alpha}\ket{\sigma,\bar{\sigma};\beta} = (\cosh\theta_{J\alpha} - \sinh\theta_{J\alpha})\bar{\sigma}_{J\alpha}\ket{\sigma,\bar{\sigma};\beta} .
\end{align}

It is clear from the above eigenstate equations, along with inverse transformations \eqref{invbog1} and \eqref{invbog2}, that states \eqref{cts} are not eigenstates of the annihilation operator $a$ of the original system. This is precisely how the expectation values of physical operators $\mcO(a,a^\dag)$ display non-trivial thermal and coherence properties simultaneously. For instance, the average number density is,
\be \bra{\sigma,\bar{\sigma};\beta}a^\dag_{J\alpha}a_{J\alpha}\ket{\sigma,\bar{\sigma};\beta} = |\sigma_{J\alpha}|^2 + \sinh^2\theta_{J\alpha}(\beta) \ee
which is indeed a sum of number densities of the coherent condensate and thermal parts.


\section{Summary and Outlook} \label{conclusion}

In this article we have presented an implementation of the thermofield dynamics formalism in the context of group field theory quantum gravity. It opens the door to using such techniques in discrete quantum gravity, thus facilitating exploration of the phase structure of quantum gravity models characterised by generalised thermodynamic parameters $\beta_\ell$, and complementing renormalization investigations in group field theory \cite{Carrozza:2016vsq,Carrozza:2017vkz,Benedetti:2014qsa} and possibly other related approaches \cite{Dittrich:2014mxa, Bahr:2014qza, Bahr:2016hwc, Bahr:2017klw}. \

Here, we have constructed finite temperature, equilibrium phases associated with a class of generalised Gibbs states in group field theory, along with identification of their non-perturbative thermal vacua. The vacua are squeezed states encoding entanglement of quantum geometric data, which in turn is expected to be a characteristic property of a physical quantum description of spacetime in general. We have introduced coherent thermal states which, in addition to carrying statistical fluctuations in a given set of observables, are also condensates of quantum geometry.  \

In group field theory, zero temperature coherent states have been used to obtain an effective description of flat, homogeneous and isotropic cosmology (flat FLRW), where certain quantum corrections arise naturally and generate a dynamical modification with respect to classical general relativity, preventing the occurrence of a big bang singularity along with cyclic solutions in general \cite{Oriti:2016qtz,Oriti:2016acw,Gielen:2016dss,Pithis:2019tvp}. Encouraged by these results, the introduction of statistical condensates, like coherent thermal states, may bring further progress to the GFT condensate cosmology program by offering a tangible and controllable way of incorporating perturbations in relevant observables. Such consideration could be crucial say for understanding the quantum gravitational origin of structure formation. One could also expect modifications during early times in the previously studied homogeneous and isotropic flat cosmology models in GFT \cite{Oriti:2016qtz,deCesare:2016rsf}, such as altering the inflation rate. This particular case of flat FLRW, in presence of thermal fluctuations, is studied in an upcoming paper \cite{KA}.

It would also be interesting to understand better the connection of our thermal vacua with other works in loop quantum gravity concerning kinematical entanglement between intertwiners (related further to discrete vector geometries) \cite{Bianchi:2016tmw,Livine:2017fgq,Baytas:2018wjd}, especially since our squeezed thermal vacua essentially encode entanglement between gauge-invariant spin network nodes, that is intertwiners, but now at a field theory level.\

The thermal vacua can be extended even further to more general two-mode squeezed vacua between tilde and non-tilde quanta, between different tilde quanta, or even between different modes of the tilde quanta. For instance, condensates of correlated quanta, like dipole condensates \cite{Gielen:2016dss}, may be directly constructed and studied in this setup. Considering correlations between different modes of the quanta, which encode quantum geometric data, might also make comparisons with studies in loop quantum gravity mentioned above \cite{Bianchi:2016tmw,Livine:2017fgq,Baytas:2018wjd} more direct. \

This setup could also prove useful for the study of quantum black holes. In group field theory for example, black holes have been modelled as generalised condensates \cite{Oriti:2018qty}, which must also possess related thermal properties. Suitably modified thermal coherent states may provide just the right type of technical structure to explore their statistical and thermal aspects further. \

Finally, by providing quite a straightforward handle on collective, quasi-particle modes in discrete quantum gravity, while still allowing for access to different inequivalent representations, this framework may bring closer the studies of microscopic theories of quantum gravity and analogue gravity models \cite{Barcelo:2005fc}. 


\begin{acknowledgments}
M.A. acknowledges the support of the project BA 4966/1-1 of the German Research Foundation (DFG) and of the Polish National Science Center OPUS 15 Grant nr 2018/29/B/ST2/01250. I.K. is grateful to Daniele Oriti for helpful discussions, and to Benjamin Bahr and University of Hamburg for the generous hospitality during visits when this project was started and partially completed. I.K. is supported by the International Max Planck Research School for Mathematical and Physical Aspects of Gravitation, Cosmology and Quantum Field Theory; and, acknowledges the hospitality of the Visiting Graduate Fellowship program at Perimeter Institute, where part of this work was carried out. Research at Perimeter Institute is supported in part by the Government of Canada through the Department of Innovation, Science and Economic Development Canada and by the Province of Ontario through the Ministry of Economic Development, Job Creation and Trade.

\end{acknowledgments}


\bibliographystyle{unsrt}
\bibliography{refvoltd}

\end{document}